\documentclass[fleqn,12pt]{wlscirep}
\usepackage[utf8]{inputenc}
\usepackage[T1]{fontenc}
\usepackage{chemformula}
\let\ce\ch
\usepackage{adjustbox}
\usepackage{lipsum}
\usepackage{mathtools}
\usepackage{physics}
\usepackage{floatrow}
\usepackage{hyperref}
\usepackage[justification=centering]{caption}
\usepackage{amsmath}
\usepackage{nccmath}
\usepackage{subcaption}
\usepackage{floatrow}
\usepackage[thinlines]{easytable}
\usepackage{cleveref}
\usepackage{gensymb}
\usepackage{booktabs}  
\usepackage{ltablex}
\usepackage{bbm}
\usepackage{graphicx}
\usepackage{subcaption}
\usepackage{lettrine}

\title{Magnetic isotope effects: a potential testing ground for quantum biology}

\author[1,2,3,*]{Hadi Zadeh-Haghighi}
\author[1,2,3,*]{Christoph Simon}

\affil[1]{Department of Physics and Astronomy, University of Calgary, Calgary, AB, T2N 1N4, Canada}
\affil[2]{Institute for Quantum Science and Technology, University of Calgary, Calgary, AB, T2N 1N4, Canada}
\affil[3]{Hotchkiss Brain Institute, University of Calgary, Calgary, AB, T2N 1N4, Canada}

\affil[
*]
{hadi.zadehhaghighi@ucalgary.ca \& csimo@ucalgary.ca}

\begin{abstract}
One possible explanation for magnetosensing in biology, such as avian magnetoreception, is based on the spin dynamics of certain chemical reactions that involve radical pairs. Radical pairs have been suggested to also play a role in anesthesia, hyperactivity, neurogenesis, circadian clock rhythm, microtubule assembly, etc. It thus seems critical to probe the credibility of such models. One way to do so is through isotope effects with different nuclear spins. Here we briefly review the papers involving spin-related isotope effects in biology. We suggest studying isotope effects can be an interesting avenue for quantum biology.

\par


\end{abstract}

\begin{document}

\maketitle

\lettrine{I}{}n atoms, the number of protons determines the element (e.g. carbon, oxygen, etc.), and the number of neutrons determines the isotope of the desired element. It has been observed that different isotopes of the element in certain chemical reactions can influence the outcomes differently. This has been shown in many chemical reactions \cite{Bigeleisen1965,Zeldovich1988,Wolfsberg2009,faure1977principles,hoefs2009stable,fry2006stable,van2011isotope,Buchachenko2001} including biological systems \cite{cook1991enzyme,Grissom1995,kohen2005isotope,buchachenko2009magnetic,Buchachenko2012,Koltover2021}. Not only do different isotopes of an element have different masses, but they can also possess different spin angular momentum, which has a magnetic property. Thus, one can consider isotope effects in (bio)chemical reactions from two distinct points of view: mass- and spin-dependency. 

Isotope effects have been reported for numerous (bio)chemical reactions \cite{Buchachenko2012,Buchachenko2013,Buchachenko2014,LBuchachenko2014,bukhvostov2014new,Buchachenko2019,Arkhangelskaya2020,Buchachenko2020,Koltover2021,Letuta2021}. Sechzer and et al. observed that administering different lithium isotopes resulted in different parenting behaviors and potentially delayed offspring development in rats \cite{Sechzer1986}. In 2020, Ettenberg co-workers \cite{Ettenberg2020}  reported that lithium isotope effect on rat's hyperactivity, where \ce{^{6}Li} produced a longer suppression of hyperactivity in an animal model of mania compared to \ce{^{7}Li}. Buchachenko et al. reported that ATP production was more than twofold in the presence of \ch{^{25} Mg} compared to \ch{^{24} Mg}. They suggested that the different nuclear spin of these isotopes was the key to these observations. The same group, in multiple studies, also observed that \ch{^{25} Mg} reduced enzymatic activity in DNA synthesis compared to \ch{^{24} Mg}, where the rate of DNA synthesis was suggested to be magnetic field-dependent \cite{Buchachenko2013d,Buchachenko2013DNA,Stovbun_2022}. They also observed isotope effects by replacing magnesium with calcium and zinc ions \cite{bukhvostov201343,Buchachenko2010a}. Li et al. observed that different xenon isotopes induced anesthesia in mice differently. In that experiment, four different xenon isotopes were used, \ce{^{129}Xe}, \ce{^{131}Xe}, \ce{^{132}Xe}, and \ce{^{134}Xe} with nuclear spins of 1/2, 3/2, 0, and 0, respectively \cite{Li2018}. They reported that isotopes of xenon with non-zero nuclear spin had lower anesthetic potency than isotopes without nuclear spin.
\par

The first mass-independent isotope effect was detected by Buchachenko and co-workers in 1976 \cite{buchachenko1976isotopic}, in which applied magnetic fields discriminated isotope effects by their nuclear spins and nuclear magnetic moments. Since then, the term "magnetic isotope effect" was introduced for such phenomena as they are controlled by electron-nuclear hyperfine coupling in the paramagnetic species. 

The sensitivity of biological systems to weak magnetic fields is an intriguing phenomenon \cite{Zadeh_Haghighi_2022}, yet incompletely understood. It is challenging to understand because the corresponding energies for such low fields are far smaller than the energies for thermal fluctuations and motions. So from a classical point of view, these effects should be washed out. But that is not the case. 
\par
One possible explanation for such effects is based on the spin dynamics of naturally occurring radical pairs, namely the radical pair mechanism \cite{Hore2016}. Spin has a magnetic property, and thus for every spin, any surrounding magnetic field from either other spins or applied magnetic field influences its state. On the other hand, spin states can determine which chemical reactions are possible, providing a mechanism for magnetic fields to influence chemical reaction products. A considerable amount of studies suggest that isotope effects in biology can be due to the spin dynamics of radical pairs in biochemical reactions.

In the context of avian magnetoreception \cite{Xu2021}, it was suggested that substituting \ce{^{17}O2} for \ce{^{17}O2} would strongly attenuate magnetosensing and also accelerate the generation of the fully oxidized state of flavin adenine dinucleotide (FAD$^{ox}$) \cite{Player2019}. Recent studies have proposed that radical pair models help explain isotope effects in xenon anesthesia \cite{Smith2021} and lithium treatment for hyperactivity \cite{Zadeh2021Li}. In these models, it is proposed that anesthesia and hyperactivity involve spin-selective electron transfer, and different isotopes of xenon and lithium influence the electron transfer process differently due to the hyperfine interaction between the xenon or lithium nuclear spin and the electron spin of the radicals, and hence possess different potency. Based on similar models, it has also been suggested that isotope effects can be tested in the role of superoxide in neurogenesis \cite{rishabh2021radical}, the effect of lithium on the circadian clock \cite{Zadeh2022CC}, and the effect of zinc on microtubule assembly \cite{ZadehHaghighi2022MT}. 
\par

It is also worth mentioning that non-mass-dependent effects or mass-independent fractionation in isotope effects have been observed with oxygen, sulfur, mercury, lead, and thallium \cite{Thiemens1983,Thiemens1999,Thiemens2001,Thiemens2006,Thiemens2012,Schauble_2007}, which are based on non-magnetic mechanisms. However, it is reported that biomolecules susceptible to oxidation by reactive oxygen species (ROS) can be protected using heavier isotopes such as \ce{^{2}H} (D, deuterium) and \ce{^{13}C} (carbon-13) \cite{Shchepinov_2007}. Moreover, in numerous studies, magnetic field effects in biology are accompanied by modulation in the ROS levels \cite{Zadeh_Haghighi_2022}. This suggests radical pairs might be involved in such ROS-related effects.

Exploring isotope effects may thus be a potential avenue to probe the radial pair mechanism hypothesis and ultimately to see whether Nature harnesses quantum physics in biology. We hope that this short article will encourage experimental experts in the field of quantum biology to test isotope effects. Furthermore, this could pave new paths for discovering new medicine and treatments.

\section*{Acknowledgment}
The authors would like to thank Michael Wieser for his valuable input. This work was supported by NSERC Discovery Grant RGPIN-2020-03945 and National Research Council CSTIP Grant QSP 022.

\bibliography{sample}

\end{document}